\documentclass[conference]{IEEEtran}
\IEEEoverridecommandlockouts
\usepackage{cite}
\usepackage{amsmath,amssymb,amsfonts}
\usepackage{algorithmic}
\usepackage{graphicx}
\usepackage{textcomp}
\usepackage[svgnames]{xcolor}
\usepackage{xspace}
\usepackage{booktabs}
\usepackage{listings}
\usepackage{hyperref}
\usepackage[capitalise]{cleveref}
\usepackage{url}
\usepackage[colorinlistoftodos,prependcaption,textsize=tiny]{todonotes}
\def\BibTeX{{\rm B\kern-.05em{\sc i\kern-.025em b}\kern-.08em
    T\kern-.1667em\lower.7ex\hbox{E}\kern-.125emX}}
\usepackage{balance}

\definecolor{diffstart}{named}{Grey}
\definecolor{diffincl}{named}{Green}
\definecolor{diffrem}{named}{Red}

\lstdefinelanguage{diff}{
	basicstyle=\ttfamily\small,
	morecomment=[f][\color{diffincl}]{+\ },
	morecomment=[f][\color{diffrem}]{-\ }
}

\newcommand{\toolname}{\emph{Code Defenders}\xspace}
\newcommand{\university}{University of Passau\xspace}

\newcommand{\summary}[2]{%
	\vspace{-0.2cm}%
	\begin{center}%
		\colorbox{gray!20}{%
			\parbox{\linewidth}{%
				\textbf{\textsf{Summary (\textit{#1})}:}~%
				#2%
			}%
		}%
	\end{center}%
}

\begin{document}

\title{An Empirical Evaluation of \\Manually Created Equivalent Mutants\\
\thanks{This work is supported by the DFG under grant \mbox{FR 2955/2-1}, ``QuestWare: Gamifying the Quest for Software Tests''.}
}

\author{\IEEEauthorblockN{Anonymous Author(s)}\vspace{2em}
}

\author{\IEEEauthorblockN{Philipp Straubinger}
	\IEEEauthorblockA{\textit{University of Passau} \\
		Passau, Germany}
	\and
	\IEEEauthorblockN{Alexander Degenhart}
	\IEEEauthorblockA{\textit{University of Passau} \\
		Passau, Germany}
	\and
	\IEEEauthorblockN{Gordon Fraser}
	\IEEEauthorblockA{\textit{University of Passau} \\
		Passau, Germany}
}

\maketitle
\begin{abstract}
  Mutation testing consists of evaluating how effective test suites
  are at detecting artificially seeded defects in the source code, and
  guiding the improvement of the test suites. Although mutation
  testing tools are increasingly adopted in practice, equivalent
  mutants, i.e., mutants that differ only in syntax but not semantics,
  hamper this process. While prior research investigated how frequently
  equivalent mutants are produced by mutation testing tools and how
  effective existing methods of detecting these equivalent mutants
  are, it remains unclear to what degree humans also create equivalent
  mutants, and how well they perform at identifying these. We
  therefore study these questions in the context of \toolname, a
  mutation testing game, in which players competitively produce
  mutants and tests. Using manual inspection as well as automated
  identification methods we establish that less than 10\% of manually
  created mutants are equivalent. Surprisingly, our findings indicate
  that a significant portion of developers struggle to accurately
  identify equivalent mutants, emphasizing the need for improved
  detection mechanisms and developer training in mutation testing.
\end{abstract}


\section{Introduction}

Mutation testing is a well-established industry practice for
identifying weaknesses in test suites and guiding test
creation~\cite{DBLP:journals/tse/PetrovicIFJ22}. It involves
introducing artificial faults (\emph{mutants}) into the source code
and assessing whether the corresponding test suites can detect them,
thus classifying mutants into killed (detected by the test suite) or
alive (undetected)~\cite{DBLP:journals/computer/DeMilloLS78}. Killed
mutants provide a quantitative assessment of the test suite quality in
terms of a mutation score (i.e., the ratio of killed mutants to
mutants overall), and surviving mutants point out to developers where
there are test gaps.
%, not all mutants are useful. Mutants introduced into defensive programming or simple data classes are often deemed useless and can be disregarded~\cite{DBLP:journals/tse/PetrovicIFJ22}.

However, not all mutants can be killed in the first place: Equivalent
mutants are semantically equivalent to the original source code, even
though they differ in syntax, such that it is impossible to create
tests that distinguish between the mutant and the original
program. Equivalent mutants skew mutation scores and are misleading
for developers, who need to manually discern whether a live mutant is
equivalent or signifies a genuine test
gap~\cite{DBLP:journals/stvr/OffuttC94}.

Contemporary mutation testing tools, such as
PIT,\footnote{\url{https://pitest.org/}} incorporate mechanisms to
identify or avoid equivalent mutants. These mechanisms often rely on
compiler
outputs~\cite{DBLP:conf/icse/PapadakisJHT15,DBLP:conf/fsen/HoushmandP17}
or specific patterns associated with equivalent
\mbox{mutants~\cite{naeem2020machine,DBLP:journals/stvr/OffuttC94}},
effectively identifying the most trivial cases. However, complex
equivalent mutants still elude automated detection and require manual
intervention by
developers~\cite{DBLP:journals/stvr/OffuttC94,DBLP:journals/stvr/SchulerZ13}.
Manual inspection of mutants, however, is time-consuming and prone to
false positives and false
negatives~\cite{DBLP:journals/stvr/SchulerZ13}.

Despite existing research on how common equivalent mutants are in
practice and how difficult they are to \mbox{detect~\cite{DBLP:journals/jss/MarsitAKLLOM21,DBLP:conf/icst/Pitts22}}, the
current body of knowledge is based on established mutation tools using
standard mutation operators to generate mutants. However, less is
known about mutants created manually, such as through mutation testing
games such as
\toolname~\cite{DBLP:conf/icst/RojasF16,DBLP:conf/sigcse/FraserGKR19,DBLP:conf/icse/RojasWCF17}.
Examining such human-generated mutants is important for several
reasons. First, developers would need to deal with such mutants when
using crowdsourcing for mutation
testing~\cite{DBLP:conf/icse/RojasWCF17}. Second, human created
mutants are an important element of software testing education, where
mutation-based assignments, exercises, and
games~\cite{DBLP:conf/sigcse/FraserGKR19} are common. Finally, a
recent trend lies in using deep learning models to mimic human edits
and produce more similar
mutants~\cite{DBLP:conf/icsm/TufanoWBPWP19,DBLP:conf/icse/TufanoKWWBPP20,degiovanni2022mubert}. In
all of these cases, a deeper understanding in the tendency of humans
to produce equivalent mutants, and their ability to detect this, is
important.

This paper aims to address the gap of knowledge on equivalence in
human generated mutants by leveraging data accumulated through several
years worth of \toolname usage in the context of a university
course. In \toolname, players can assume the role of an attacker,
introducing mutants, or a defender, creating tests to detect
mutants. To cope with equivalent mutants, the game includes dedicated
features focusing on equivalence, such as equivalence duels, where one
player can claim a mutant as equivalent, and the mutant creator must
either disprove the equivalence or accept the claim.
Using the resulting dataset of mutants and tests for a set of Java
classes, we investigate how many mutants can be killed by existing
tests, how many of the remaining mutants can be detected as equivalent
by automated equivalence detection techniques, and how many are
actually equivalent. 
In detail, the contributions of this paper are as follows:

\begin{enumerate}
	\item We provide an extensive dataset comprising 18,000 manually
	written mutants and 11,000 corresponding tests in ten different Java classes.
	\item We evaluate how well existing automated methods to identify
	trivial equivalent mutants perform on manually created equivalent
	mutants.
	\item We manually classify a substantial number of equivalent mutants,
	and assess the ability of players of \toolname to
	identify and classify equivalent mutants.
\end{enumerate}

The results reveal that less than 10\% of the manually created mutants
are equivalent, and nearly two-thirds of players were unable to
accurately identify manually created equivalent mutants. This may have
implications beyond education, questioning the validity of equivalent 
mutation classifications by humans in general.

%%% Local Variables:
%%% mode: latex
%%% TeX-master: "../main"
%%% End:

\section{Background}

\subsection{Mutation Testing}

Mutation testing describes the process of seeding artificial defects
(mutants) into source code to identify weaknesses in existing
test suites.  The use of small, artificial defects is based on the
\textit{Coupling Effect} and \textit{Competent Programmer}
hypotheses~\cite{DBLP:conf/afips/BuddLDS78}. The former suggests that
tests revealing simple errors are effective in also uncovering more
complex errors, while the latter posits that programmers create nearly
correct programs with small deviations from the correct
program~\cite{DBLP:journals/computer/DeMilloLS78}.  Mutants can be
classified based on the result of executing available tests against
them: A failing test indicates a killed mutant, while mutants survive
if all tests pass. Live mutants imply potential test suite
deficiencies, and developers can use this information to strengthen
their test suites. The ratio of killed to generated mutants represents
the mutation score, and serves as a metric correlated with a test
suite's fault-detection ability, surpassing metrics like code
coverage~\cite{walsh1985measure}.

While the theoretical aim would be to achieve a mutation score of
100\%, and thus having a strong test suite able to detect all mutants,
this goal is usually not achievable, as mutants may be semantically
equivalent despite their syntactical differences. An equivalent mutant
cannot possibly be killed by a test.

\subsection{Equivalent Mutants}

Equivalent mutants are mutated versions of the original code that,
despite their altered structure, produce the same output as the
unmutated code when subjected to a set of
tests~\cite{DBLP:journals/stvr/OffuttC94}. These mutants introduce
functionally equivalent changes, making them challenging to
distinguish from the original code solely based on traditional testing
methods. Since finding equivalent mutants is an undecidable
problem~\cite{DBLP:journals/acta/BuddA82}, it is mostly done using
heuristics and partial solutions~\cite{DBLP:journals/stvr/OffuttC94}.

Detecting equivalent mutants involves addressing a binary
classification task with potential errors: False negatives occur when
an equivalent mutant is erroneously labeled as non-equivalent, while
false positives represent non-equivalent mutants inaccurately marked
as equivalent. In mutation testing, false positives are considered
more critical as they lead to the exclusion of potentially important
killable mutants~\cite{DBLP:journals/stvr/OffuttC94}.

Manual analysis is often necessary to identify equivalent mutants but
may require significant time investment and is prone to false
positives. For example, Schuler and
Zeller~\cite{DBLP:journals/stvr/SchulerZ13} report that manual
classification took 15 minutes per mutant. Yao et
al.~\cite{yao2014study} reported 6 person-months of manual analysis
effort dedicated entirely to classifying 1230 mutants.
Automated techniques for detecting equivalent mutants are thus
desirable, and fall into two main
categories~\cite{DBLP:conf/icse/PapadakisJHT15}: \textit{Detect
	Approaches}, which directly determine whether a mutant is
equivalent, and \textit{Reduce Approaches}, which provide an order from less
likely to more likely equivalent mutants.
%but does not guarantee
%freedom from false positive errors.

Recently, Trivial Compiler Equivalence
(TCE~\cite{DBLP:conf/icse/PapadakisJHT15}) and its extended version
TCE+~\cite{DBLP:conf/fsen/HoushmandP17} have been demonstrated to be
effective at identifying equivalent mutants. TCE identifies equivalent
mutants by comparing output files after the compilation of the source
code, assuming that equivalent mutants will produce identical compiled
outputs. TCE+ builds upon this approach by incorporating an additional
optimization step post-compilation, enhancing its effectiveness,
especially in languages like Java, where optimization occurs at
runtime. Unlike TCE, TCE+ compares output files after the optimization
step.

The increasing adoption of machine learning approaches in software
engineering has also resulted in predictive approaches not only to
speed up mutation testing~\cite{zhang2016predictive}, but also to
classify equivalent
mutants~\cite{naeem2020machine,peacock2021automatic,brito2020preliminary,jammalamadaka2022equivalent}.
%This method
%explores the use of constraint-based testing (CBT) theory to identify
%equivalent mutants, utilizing metrics such as reachability, necessity,
%and sufficiency to construct a binary classification model.
Such machine learning approaches require datasets of equivalent
mutants such as MutantBench~\cite{van2021mutantbench}, but since these
datasets require substantial manual labelling effort there are efforts
to use automation to automatically augment equivalent mutant
datasets~\cite{chung2022augmenting}.

\subsection{Code Defenders}

\toolname is a web application that gamifies Mutation Testing,
where players have to competitively create mutants and writing tests to kill
them~\cite{DBLP:conf/icst/RojasF16}. \toolname was devised as a
crowdsourcing platform to elicit strong mutants and tests, and also
serves as a valuable tool for teaching software testing, addressing
the common perception among students is that testing is more tedious than
software development itself~\cite{DBLP:conf/icse/ElbaumPDJ07}. By
incorporating mutation testing, \toolname familiarizes aspiring
developers with testing concepts, potentially increasing the practical
application of mutation testing in real-world
scenarios~\cite{DBLP:conf/fie/OliveiraOCD15}. \toolname has been
positively received by students, enhancing their testing skills during
a software testing course~\cite{DBLP:conf/sigcse/FraserGKR19}. It is
publicly accessible,\footnote{\url{https://code-defenders.org}}
open-source,\footnote{\url{https://github.com/CodeDefenders/CodeDefenders}}
and therefore used by different universities globally.

% \begin{lstlisting}[float, caption=Test template of \toolname, captionpos=b, label=lst:test, linewidth=\columnwidth, basicstyle=\footnotesize\ttfamily, columns=fullflexible, language=Java]
% // package declaration if necessary

% import org.junit.Test;
% // additional imports if necessary

% public class Test<CUT Name> {

% 	@Test(timeout = 4000)
% 	public void test() throws Throwable {
% 		// test here!
% 	}
% }
% \end{lstlisting}

\subsubsection{Game Modes}

\toolname currently offers three game modes: The \emph{Puzzle Mode} is
a single-player experience where players solve predefined tasks using
mutants/tests. The \emph{Battleground Mode} represents the default
multiplayer experience, where players are divided into teams of
attackers and defenders, competing over a Java class under
test. Finally, the \emph{Melee Mode} is a multiplayer option where all
players compete against others, each player taking on both attacker
and defender roles simultaneously.

\subsubsection{Testing}

Depending on the game mode (i.e., \textit{easy} vs. \textit{hard}),
defenders can either see only the locations of mutants in the game, or
also their diff, and then have to create tests to kill these mutants.
Tests are created in a web-based user interface
%using the template
%shown in \cref{lst:test}, where players are only allowed to modify
%lines within the \texttt{test()} method. Tests
and may use various common
test libraries such as
JUnit5,\footnote{\url{https://junit.org/junit5/}}
Hamcrest,\footnote{\url{https://hamcrest.org/JavaHamcrest/}} and
Google Truth.\footnote{\url{https://github.com/google/truth}} When
tests are submitted, they undergo compilation and validation checks
before acceptance into the game. The validation ensures tests are
non-flaky, deterministic, concentrate on testing a limited set of
functionalities, and pass on the original class under test.  Depending
on the game mode there are certain restrictions on what code is
permitted in the tests to ensure fairness and clean tests. For
example, \toolname checks submitted tests to ensure they do not
contain loops, calls to \texttt{System.*}, additional methods,
conditionals, or exceed a configurable maximum number of test
assertions.

\begin{figure*}
	\centering
	\includegraphics[width=\linewidth]{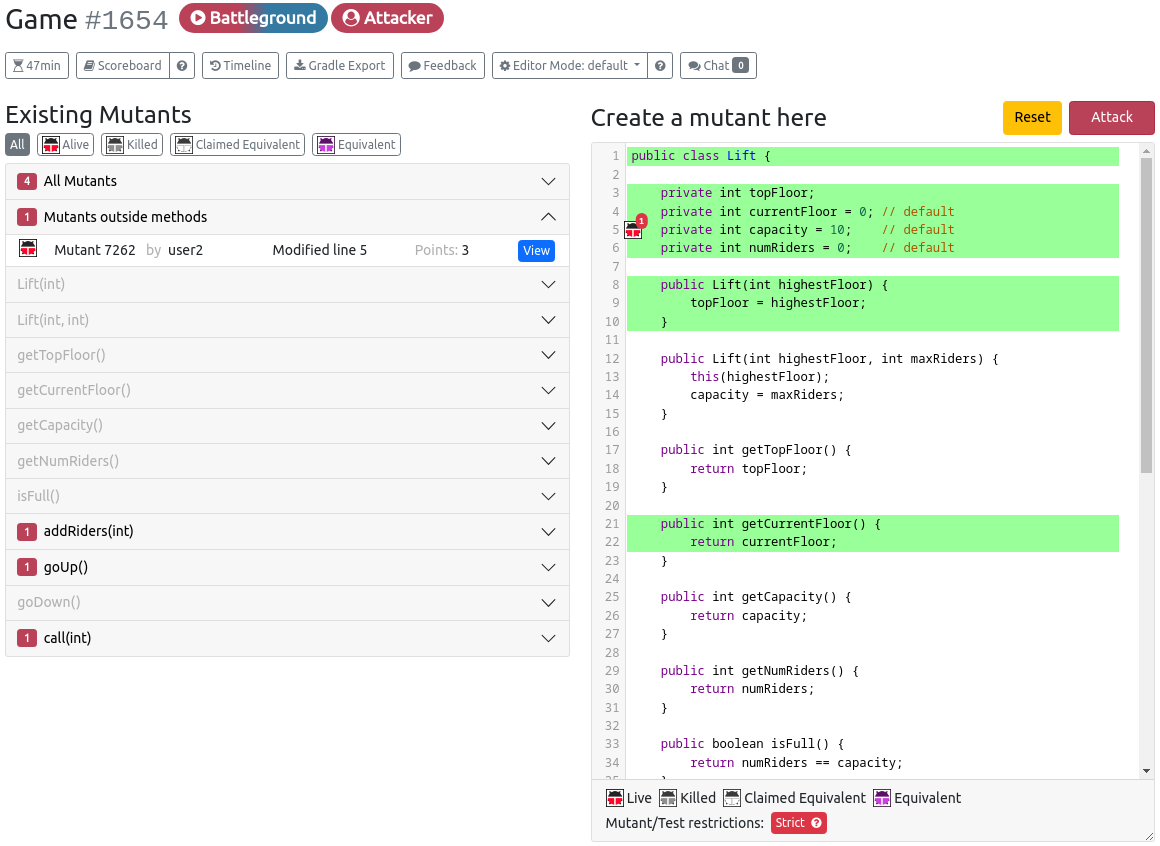}
	\caption{Attacker view of \toolname}
	\label{fig:attacker}
\end{figure*}

\subsubsection{Mutation}

Attackers create mutants by editing the CUT and submitting their
modified versions (\cref{fig:attacker}). \toolname first validates
these mutants based on a configurable \textit{Mutant Validation
	Level}, which can be categorized as \textit{strict},
\textit{moderate}, or \textit{relaxed}. These levels try to prevent
both equivalent and unfair mutants, e.g., relying on random values.
In the \textit{relaxed} level, there are minimal restrictions, only
disallowing calls to \texttt{System.*} and \texttt{Random}.  The
\textit{moderate} level targets mutants that might be hard to kill but
offer no value for testers. Restrictions include modifying comments,
adding additional logical operators and control structures, as well as
ternary operators.  The \textit{strict} level prohibits adding bitwise
operators, using reflection, and modifying signatures. \toolname also
performs basic equivalence detection by stripping whitespaces and
comparing mutant and original CUT. This prevents intentional or
accidental submission of equivalent mutants (e.g., submitting after
adding only a new line).

In addition to the validation of the restrictions, a hash is computed
based on the whitespace-stripped mutant code. This hash is used to
identify duplicate mutants within a game. If a mutant with the same
hash already exists, the newly submitted mutant is rejected. This
serves as a basic duplicate detection approach, preventing players
from submitting identical mutants multiple times by accident or for
point farming.

\begin{figure*}
	\centering
	\includegraphics[width=\linewidth]{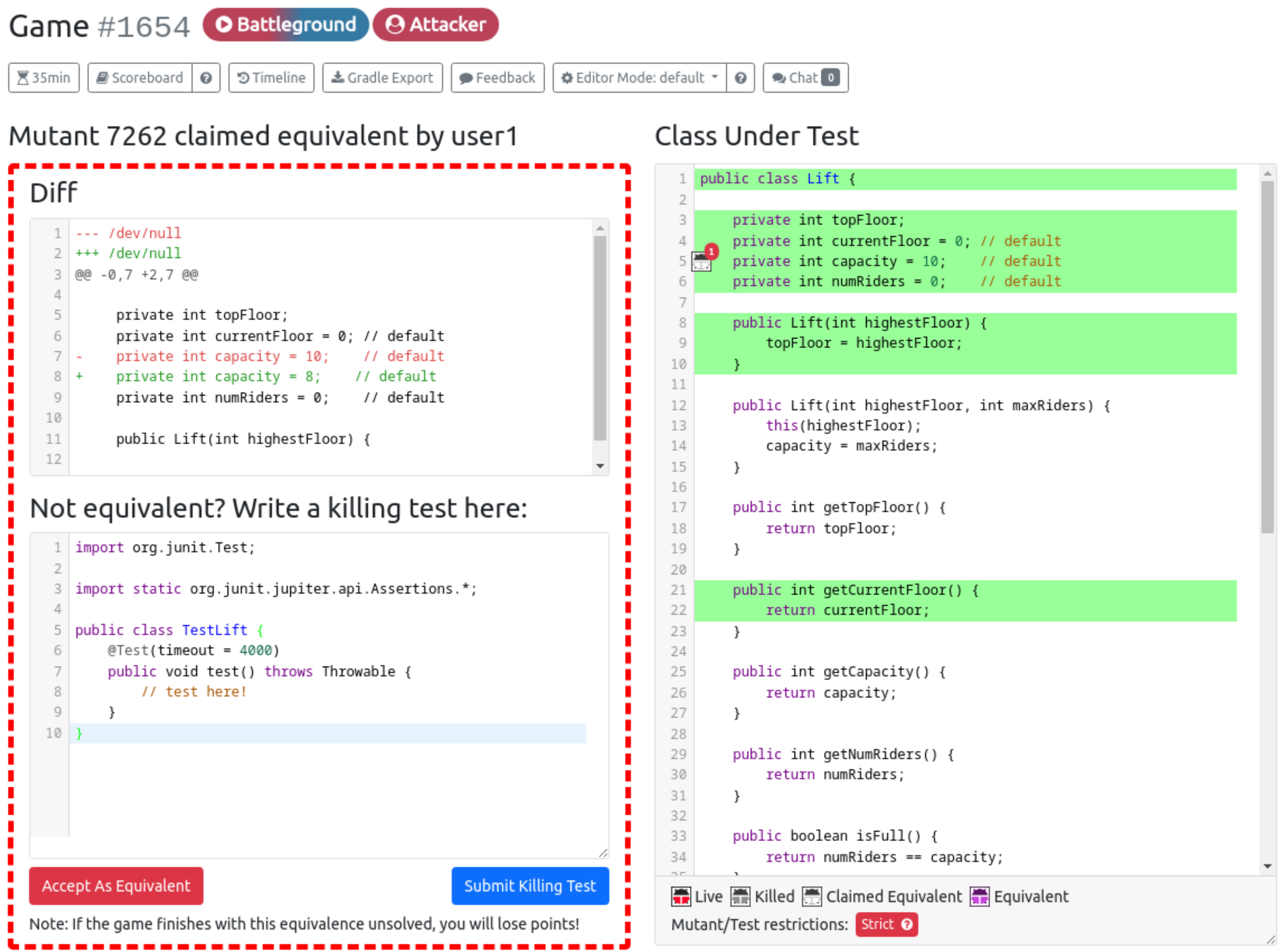}
	\caption{Equivalence duel during a game of \toolname}
	\label{fig:duel}
\end{figure*}

\subsubsection{Intent Collection}

While defenders always need to reason about code behavior and existing
mutants when creating tests, attackers may, in particular in earlier
phases of the game where the coverage achieved by the defenders is
still low, arbitrarily mutate code without deeper thought. While this
is a strategy that likely backfires later in the game, it is
particularly undesirable in an educational
context~\cite{DBLP:conf/sigcse/FraserGKR19}.  Therefore, a
configurable feature aiming to force players to think more deeply
about their actions before submitting them is \emph{intent
	collection}: If enabled, players are required to provide additional
metadata when submitting mutants or tests. In particular, attackers
have to specify whether they intended to create a \textit{killable} or
\textit{equivalent} mutant, or they can choose \textit{don't know} if
they are uncertain about the mutant's equivalence.  Defenders must
select a line in the CUT they intend to target when intent collection
is activated.

\subsubsection{Equivalence Duels}

Regardless of whether created on purpose or by accident, equivalent
mutants are a common aspect of games, and therefore integrated using
the concept of \emph{Equivalence Duels}. After defenders have
attempted to kill a mutant and managed to create at least one test
that covers it without killing it, they challenge the attacker
who created the mutant to such a duel. \toolname can also be
configured to automatically trigger these duels if a mutant has been
covered by a specified number of tests without being killed.

When challenged to an equivalence duel for a mutant they created, the
attacker is temporarily blocked from submitting more mutants until the
duel is resolved (\cref{fig:duel}). The duel can be resolved by the
attacker by submitting a valid test that kills the mutant, thus
proving its non-equivalence and earning the attacker a win. Submitting
a valid test that does not kill the mutant results in the defender
suspecting the mutant to be equivalent winning the
duel. Alternatively, the attacker can also accept the mutant as
equivalent, leading the defender to win the duel and assuming the
mutant's equivalence. To incentivize attackers to invest time in
thoughtful testing, losing an equivalence duel results in the loss of
all accumulated points for the mutant, while winning the duel earns
the attacker an additional point. Independently of the game difficulty
setting, attackers always get to see the diff of the mutant that is
part of the duel, since they created it in the first place.

%%% Local Variables:
%%% mode: latex
%%% TeX-master: "../main"
%%% End:

\section{Evaluation}

To understand the role of equivalent mutants within \toolname, we aim
to answer the following research questions:

\begin{itemize}
	\item \textit{\textbf{RQ 1}: How well does TCE(+) perform on manually written mutants?}
	\item \textit{\textbf{RQ 2}: How many equivalent mutants do players of \toolname create?}
	\item \textit{\textbf{RQ 3}: How well do players perform at detecting (non-) equivalent mutants?}
\end{itemize}

\subsection{Dataset}

\begin{table}[]
	\caption{Overview of classes used for analysis}
	\label{tab:classes}
	\centering
	\begin{tabular}{lrr}
		\toprule
		\textbf{CUT Alias} & \textbf{Years played} & \textbf{Number of Games} \\ \midrule
		ByteVector         & 1                     & 8                        \\
		Complex\_V1        & 2                     & 23                       \\
		Complex\_V2        & 1                     & 20                       \\
		Document           & 3                     & 24                       \\
		HSLColor           & 1                     & 8                        \\
		IntHashMap         & 5                     & 40                       \\
		Lift               & 3                     & 37                       \\
		Options            & 3                     & 24                       \\
		Rational           & 2                     & 36                       \\
		SparseIntArray     & 5                     & 43                       \\ \bottomrule
	\end{tabular}
\end{table}

Our dataset encompasses information gathered from sessions using
\toolname during Software Testing lectures at \university over the
last five years (2018--2022).
%Each year, a dedicated \toolname
%instance was set up, followed by the creation of a backup containing a
%full database dump and data archive.
From this data, we extracted the source code for the Classes Under
Test (CUTs), along with details about the number and types of games,
mutants, and tests associated with each game. Our focus is on
battleground games, the most established game type used consistently
across all years.

Throughout these years, various CUTs were utilized in \toolname
sessions, predominantly chosen from an available
pool~\cite{DBLP:conf/sigcse/FraserGKR19}. To ensure a sufficient
number of mutants and tests for each CUT, we included all CUTs played
in at least two years. However, changes to attribute visibilities and
additional getters in the \texttt{Complex} class led to two different
versions of this class, which we count as two distinct CUTs in the
dataset. The \toolname instances not only featured games played
seriously but also those created exclusively for testing or
demonstration purposes. Consequently, we excluded games with fewer
than 15 submitted mutants and tests, respectively.
This process resulted in a total of 10 CUTs (\cref{tab:classes}).
%,
%forming a subset of those originally employed in the work by Fraser et
%al.~\cite{DBLP:conf/sigcse/FraserGKR19}.

To prepare the dataset for further evaluation, we compiled and
executed all extracted tests against the corresponding CUT,
eliminating any tests that either failed to compile or did not pass
against the unchanged CUT. In a subsequent step, we compiled the
mutants, discarding those that did not compile. The tests from the
preceding step were then executed against each mutant. Any mutants for
which at least one test fails were then classified as killable, as
they cannot be equivalent.

\begin{table}[]
	\caption{Killed mutants}
	\label{tab:killmutants}
	\centering
	\resizebox{\linewidth}{!}{%
		\begin{tabular}{lrrrrrr}
			\toprule
			\textbf{CUT Alias} & \textbf{Total} & \textbf{Kill.} & \textbf{Alive} & \textbf{Kill. \%} & \textbf{Alive \%} & \textbf{Tests} \\ \midrule
			ByteVector         & 608            & 477            & 131            & 78.45\%             & 21.55\%             & 236            \\
			Complex\_V1        & 1447           & 1341           & 106            & 92.67\%             & 7.33\%              & 517            \\
			Complex\_V2        & 1374           & 1281           & 93             & 93.23\%             & 6.77\%              & 797            \\
			Document           & 1791           & 1651           & 140            & 92.18\%             & 7.82\%              & 1009           \\
			HSLColor           & 739            & 553            & 186            & 74.83\%             & 25.17\%             & 342            \\
			IntHashMap         & 3472           & 3253           & 219            & 93.69\%             & 6.31\%              & 2935           \\
			Lift               & 1862           & 1684           & 178            & 90.44\%             & 9.56\%              & 1254           \\
			Options            & 1845           & 1758           & 87             & 95.28\%             & 4.72\%              & 684            \\
			Rational           & 1383           & 1205           & 178            & 87.13\%             & 12.87\%             & 681            \\
			SparseIntArray     & 3702           & 3392           & 310            & 91.63\%             & 8.37\%              & 2806           \\ \midrule
			Total			   & 18223			& 16595			 & 1628			  & 91.07\%			  & 8.93\%			  & 11261 \\ \bottomrule
		\end{tabular}
	}
\end{table}

\cref{tab:killmutants} presents an overview of both the killed and
alive mutants after running the tests against them. It also includes
the total number of mutants and tests separated by CUT. Notably, more
than 90\% of all valid mutants in the datasets were killed by
tests. For the majority of CUTs, the ratio of killed mutants exceeds
90\%. However, \texttt{ByteVector} and \texttt{HSLColor} stand out as
apparent outliers. Several factors could contribute to their lower
percentage of killed mutants. One possibility is that fewer games were
played with these two classes, resulting in fewer accumulated tests,
particularly for corner cases (about 200 to 300 tests for
\texttt{ByteVector} and \texttt{HSLColor} compared to more than 500
for all other CUTs). Another factor could be that the games involving
these two classes utilized the \textit{Intention Collection} feature,
potentially influencing how many equivalent mutants attackers create.

\subsection{Analysis Procedure}

\subsubsection{RQ 1: How well does TCE(+) perform on manually written mutants} \label{sec:rq3}

The first research question aims to evaluate how well the
state-of-the-art approaches for identifying equivalent mutants,
TCE/TCE+, performs at identifying equivalent mutants created in
\toolname games. Trivial Compiler Equivalence
(TCE)~\cite{DBLP:conf/icse/PapadakisJHT15} identifies mutants as
equivalent if their compiled files match the compilation output of the
original CUT. However, since Java's compilation involves minimal
optimization, particularly compared to languages like C or Fortran,
TCE may not be as effective. To address this, we also utilize TCE+, an
extension of TCE that incorporates an optimization step after
compilation, utilizing the optimized class files for equivalence
detection~\cite{DBLP:conf/fsen/HoushmandP17}.

The required files for TCE are obtained by compiling the sources,
while for TCE+ detection, the class files must undergo an optimization
after compilation. Following the approach in the original TCE+
paper~\cite{DBLP:conf/fsen/HoushmandP17}, we employ
ProGuard,\footnote{\url{https://github.com/Guardsquare/proguard}} an
open-source shrinker and optimizer designed for Java, primarily aimed
at Android apps. Our
configuration of ProGuard retains all classes, methods, and attributes regardless
of their access modifiers, a crucial consideration as tests may employ
reflection to access attributes or methods.

To ensure that ProGuard optimization retains all components used by
the tests, the initial step involves optimizing the CUTs, followed by
executing the test suites against the optimized versions to confirm
the success of all tests. Subsequently, the mutants undergo
optimization as well. To assess whether the class files of a mutant
(whether normal or optimized) match those of the CUT, we generated
\texttt{SHA256} hashes for all \texttt{.class} files and verified
whether the checksums of all mutant files correspond to those of the
CUT.

This leaves us with a set of mutants that are neither killed by tests,
nor flagged as equivalent by TCE or TCE+. Next, we manually examined a random sample constituting 20\% of the remaining mutants in each subgroup: those not eliminated by a test in a duel, mutants marked as equivalent in a duel, and mutants neither involved in a duel nor eliminated. This process yielded an estimated ratio of equivalent mutants in the initial dataset, which we could then compare with the number of equivalent mutants identified by TCE and TCE+.

\subsubsection{RQ 2: How many equivalent mutants do players of \toolname create?} 

The overall dataset combines mutants from multiple games for multiple
classes.  To understand player behavior in games and answer this
research question, we reuse the data containing both the automatically
and manually tagged equivalent mutants gathered from RQ1.

\subsubsection{RQ 3: How well do players perform at detecting \mbox{(non-)}~equivalent mutants}

\begin{table}[]
	\caption{Number of (resolved) duels}
	\label{tab:totalduels}
	\centering
	\begin{tabular}{lrrr}
		\toprule
		CUT Alias      & Duels & Mutants killed outside & Resolved duels \\ \midrule
		ByteVector     & 66    & 6                      & 34             \\
		Complex\_V1    & 36    & 2                      & 26             \\
		Complex\_V2    & 59    & 0                      & 43             \\
		Document       & 276   & 17                     & 186            \\
		HSLColor       & 190   & 17                     & 129            \\
		IntHashMap     & 1067  & 113                    & 822            \\
		Lift           & 215   & 8                      & 137            \\
		Options        & 196   & 25                     & 83             \\
		Rational       & 129   & 7                      & 69             \\
		SparseIntArray & 693   & 41                     & 554            \\ \midrule
		Total          & 2927  & 236                    & 2082           \\ \bottomrule
	\end{tabular}
\end{table}

To answer this research question, we consider the two mechanisms
intended to address equivalent mutants in \toolname: 
First, we extract equivalence duels from the database, including the
current state of the mutant under consideration. We compare the ratio
of equivalent mutants with and without equivalence duels and further
analyze mutants with automatically triggered versus manually triggered
equivalence duels. Additionally, we examine player actions in
equivalence duels based on whether the duel subject is an equivalent
mutant or not. For equivalent mutants, this involves whether the
player accepted the mutant as equivalent or submitted a killing
test. For non-equivalent mutants, we also investigated whether the
player accepted the mutant as equivalent, submitted a killing test,
and assessed whether that test successfully killed the mutant.

To maintain data consistency and meaningfulness, six mutants from a single \texttt{Document} CUT game were excluded from further analysis. In these cases, the attacker provided a test that killed the mutant, yet these mutants were not killed in the analysis, with one even identified as equivalent. This discrepancy may have arisen from mishandling an edge case or a temporary system problem during that specific game.

There are several instances where mutants included in equivalence duels are killed outside of their duel, which occurs when a defender submits a new test after a duel has been triggered (\cref{tab:totalduels}).
This can result in the mutant being killed by the defender, causing issues with the duel resolution process.
\cref{tab:totalduels} shows that this affects at least 5\% of all equivalence duels for nearly all CUTs. To conduct further analysis, we exclude these mutants from the study as their deaths outside of the duel prevent a normal resolution status from being available. Additionally, we also excluded duels that were not resolved by the end of the game, where the resolving rate ranged between 49\% and 86\% for the different CUTs.

Second, we extract intention information and correlate it with the
data collected in \cref{sec:rq3} regarding whether mutants were killed
or identified as equivalent. For the remaining unknown mutants, we
apply the same sampling strategy as in \cref{sec:rq3} to manually
investigate 20\% per subgroup (intended equivalent, intended not equivalent, and not provided), providing an estimate for all mutants. We then
analyze how many mutants have a correct, incorrect, or unknown intent,
and explore whether players performed equally well in classifying 
equivalent and non-equivalent mutants or if they were more adept at
identifying one over the other.

\subsection{Threats to Validity}

\paragraph{Threats to external validity} Potential variations in
surrounding conditions during game sessions and the dataset's
specificity to the Code Defenders session at the university may limit
the generalizability of results to other contexts or CUTs. In
particular intent information for mutants is available only for two
CUTs (\texttt{ByteVector} and \texttt{HSLColor}). Changes in
dependencies, Java versions, and \toolname over the years may further
limit the external validity of the findings. Involving only students 
in generating mutants and conducting tests could reduce the 
applicability of the findings and might yield different outcomes 
compared to professionals working in industry settings. However, 
it is worth noting that the students who took part in this study 
were nearing the completion of their Bachelor's degrees, 
implying they already possessed a certain level of knowledge 
and experience in testing.

\paragraph{Threats to internal validity}
%The structure of the game sessions evolved, with changes in the number of games played, the mutants generated, the tests administered, and the duration of each game. These variations might lead to different outcomes if the experiment were repeated in the future. However, since we took the results from various games and setups, the general findings should remain consistent.

%\paragraph{Threats to construct validity}
The manual analysis of mutants introduces the potential for
misclassification, which we tried to minimize using a combination of
large test suites dedicated to detecting mutants and automated
detection methods. Our sampling procedure for classifying
subpopulations of mutants may result in a bias, which we tried to
mitigate using a large sample with substantial manual classification
effort.

%%% Local Variables:
%%% mode: latex
%%% TeX-master: "../main"
%%% End:

\section{Results}

\subsection{RQ 1: How many equivalent mutants are detected automatically?}

\begin{table}[t]
	\caption{Automatically detected mutants based on alive mutants}
	\label{tab:automutants}
	\centering
%	\resizebox{\linewidth}{!}{%
		\begin{tabular}{lrrrr}
			\toprule
			\textbf{CUT Alias} & \textbf{Alive} & \multicolumn{2}{c}{\textbf{Detected}} & \textbf{Unknown} \\ 
                        & & \textbf{Total} & \textbf{\%} & \\ \midrule
                        ByteVector         & 131            & 26                       & 19.85\%                       & 105                        \\
			Complex\_V1        & 106            & 39                       & 36.79\%                       & 67                         \\
			Complex\_V2        & 93             & 28                       & 30.11\%                       & 65                         \\
			Document           & 140            & 22                       & 15.71\%                       & 118                        \\
			HSLColor           & 186            & 46                       & 24.73\%                       & 140                        \\
			IntHashMap         & 219            & 63                       & 28.77\%                       & 156                        \\
			Lift               & 178            & 56                       & 31.46\%                       & 122                        \\
			Options            & 87             & 3                        & 3.45\%                        & 83                         \\
			Rational           & 178            & 60                       & 33.71\%                       & 118                        \\
			SparseIntArray     & 310            & 91                       & 29.35\%                       & 219                        \\ \midrule
                  Total			   & 1628			& 434					   & 26.66\%						 & 1193 \\
                  \bottomrule
		\end{tabular}
%	}
\end{table}

\Cref{tab:automutants} provides an overview of the live mutants before
automatic detection, excluding mutants killed by tests contained in
the dataset, and the number of mutants automatically identified as
equivalent by either TCE or TCE+ alongside the remaining unknown
mutants, which were neither automatically killed nor flagged as
equivalent.

Notably, of the approximately 1600 mutants remaining after removing
all killed mutants, more than a fourth were detected as
equivalent. The ratio of detected equivalent mutants varies across
CUTs, ranging from around 3\% to nearly 37\%. No apparent reasons
explain the variations in the ratio of detected equivalent mutants
among the remaining live mutants. One minor observation is that the
\texttt{Options} CUT, with the lowest ratio of detected equivalent
mutants (i.e., 3\%), also had the highest ratio of killed mutants
(i.e., 95\%), although this is likely coincidental, as related
comparisons do not align (i.e., the highest ratio of detected
equivalent mutants do not correspond to the lowest ratio of killed
mutants).

\begin{figure}[t]
	\includegraphics[width=\linewidth]{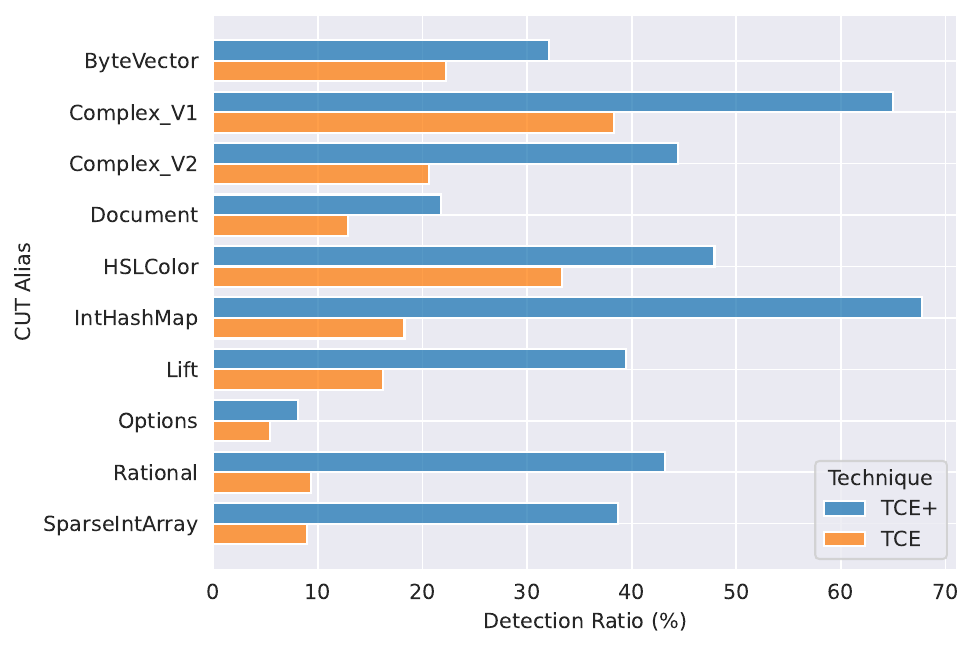}
	\centering
	\caption{Equivalent Mutant Detection Ratios for the TCE and TCE+ techniques}
	\label{fig:autodetection}
\end{figure}

\begin{lstlisting}[float, language=diff, float, caption=Equivalent mutant in \texttt{Document} found by TCE+\, but not by TCE, captionpos=b, label=lst:tce, linewidth=\columnwidth]
   while (it.hasNext()) {
-    IndexableField field = it.next();
+    IndexableField field;
+    field = it.next();
     if (field.name().equals(name)) {
       it.remove();
\end{lstlisting}

The manual classification of 22.5\% (268) of the remaining mutants not
identified by TCE/TCE+ allows us to estimate the total number of
equivalent mutants, and the detection ratio of both techniques. TCE+
achieves an equivalent mutant detection ratio ranging from around 32\%
to 48\% in most cases (see \cref{fig:autodetection}), with the best
detection ratio exceeding 65\%. On the other hand, TCE detects a
maximum of 38\% of all equivalent mutants, with detection ratios
between 9\% and 22\%. Overall, TCE detected 16.7\% of all 
equivalent mutants, whereas TCE+ managed to detect 41.5\%.

All mutants identified as equivalent by TCE were also detected by
TCE+, and TCE+ consistently outperforms TCE across all CUTs
(\cref{fig:autodetection}). The degree of improvement with TCE+
compared to TCE varies among different CUTs. For some classes (e.g.,
\texttt{Rational} and \texttt{SparseIntArray}), TCE+ identifies
approximately four times as many equivalent mutants as TCE, while for
others (e.g., \texttt{ByteVector}, \texttt{HSLColor}, and
\texttt{Options}), the additional optimization step reveals only
around 50\% more mutants. An example of a mutant detected by TCE+ but
not by TCE is illustrated in \cref{lst:tce}. In this case, a field's
declaration and initialization are separated into two lines, a
similarity that TCE+ can recognize because it results in the same Java
bytecode after optimization.

\summary{RQ 1}{TCE and TCE+ successfully identified more than a quarter of the remaining alive mutants as equivalent. TCE+ consistently outperformed TCE in detecting equivalent mutants across different CUTs, with a detection rate of 41.5\%, while TCE only achieved 16.7\%.}

\subsection{RQ 2: How many equivalent mutants do players of \toolname create?}

\begin{table}[]
	\caption{Equivalent Mutant Ratios per CUT for all mutants}
	\label{tab:equimutants}
	\centering
	%	\resizebox{\linewidth}{!}{%
	\begin{tabular}{lrrr}
		\toprule
		\textbf{CUT Alias} & \textbf{Detected} & \textbf{Estimated} & \textbf{Total} \\ \midrule
		ByteVector         & 4.28\%                        & 9.05\%                       & 13.32\%                    \\
		Complex\_V1        & 2.70\%                        & 1.42\%                       & 4.12\%                     \\
		Complex\_V2        & 2.04\%                        & 2.55\%                       & 4.59\%                     \\
		Document           & 1.23\%                        & 4.39\%                       & 5.62\%                     \\
		HSLColor           & 6.22\%                        & 6.77\%                       & 12.99\%                    \\
		IntHashMap         & 1.81\%                        & 0.87\%                       & 2.68\%                     \\
		Lift               & 3.01\%                        & 4.64\%                       & 7.65\%                     \\
		Options            & 0.16\%                        & 1.85\%                       & 2.01\%                     \\
		Rational           & 4.34\%                        & 5.69\%                       & 10.03\%                    \\
		SparseIntArray     & 2.46\%                        & 3.90\%                       & 6.36\%                     \\ \midrule
		Total			   & 2.38\%						& 3.36\%						& 5.75\%						\\ \bottomrule
	\end{tabular}
	%	}
\end{table}

\begin{table*}[h!]
	\caption{Number of resolved duels and their outcomes with EM = equivalent mutant}
	\label{tab:duels}
	\centering
	\resizebox{\linewidth}{!}{%
		\begin{tabular}{lrrrrrrr}
			\toprule
			CUT Alias      & Resolved duels & Resolved duels / game & Mutant killed by test & EM survived test & Non EM survived test & Correctly accepted & Wrongly accepted \\ \midrule
			ByteVector     & 34             & 4.25                    & 8                      & 4                    & 3                        & 17                 & 5                \\
			Complex\_V1    & 26             & 1.13                    & 3                      & 4                    & 1                        & 9                  & 10               \\
			Complex\_V2    & 43             & 2.15                    & 20                     & 8                    & 5                        & 6                  & 9                \\
			Document       & 186            & 7.75                    & 94                     & 28                   & 21                       & 41                 & 23               \\
			HSLColor       & 129            & 16.13                   & 60                     & 23                   & 17                       & 38                 & 8                \\
			IntHashMap     & 822            & 20.55                   & 209                    & 179                  & 162                      & 55                 & 379              \\
			Lift           & 137            & 3.70                    & 47                     & 27                   & 13                       & 39                 & 23               \\
			Options        & 83             & 3.46                    & 29                     & 21                   & 17                       & 11                 & 22               \\
			Rational       & 69             & 1.92                    & 13                     & 15                   & 7                        & 33                 & 7                \\
			SparseIntArray & 554            & 12.88                   & 157                    & 132                  & 88                       & 93                 & 172              \\ \midrule
			Total          & 2082           & 7.92                    & 640                    & 441                  & 334                      & 342                & 658              \\ \bottomrule
		\end{tabular}
	}
\end{table*}

\begin{table}[h!]
	\caption{Results of the Mutant Intentions given by the players, and their equivalence status}
	\label{tab:mutantintention}
	\centering
	\resizebox{\linewidth}{!}{%
		\begin{tabular}{lrrrr}
			\toprule
			\textbf{CUT Alias} & \textbf{Total} & \textbf{Correct \%} & \textbf{Incorrect \%} & \textbf{Not provided \%} \\ \midrule
			\multicolumn{5}{l}{\textit{Intention given by the players}}                                                                                                                      \\ \midrule
			ByteVector                          & 608                             & 88.32\%                               & 4.11\%                                  & 7.57\%                                     \\
			HSLColor                            & 739                             & 88.36\%                               & 6.22\%                                  & 5.41\%                                     \\ \midrule
			\multicolumn{5}{l}{\textit{When the intention was not equivalent}}                                                                                                               \\ \midrule
			ByteVector                          & 527                             & 92.18\%                               & 1.53\%                                  & 6.3\%                                      \\
			HSLColor                            & 643                             & 93.07\%                               & 1.23\%                                  & 5.7\%                                      \\ \midrule
			\multicolumn{5}{l}{\textit{When the intention was equivalent}}                                                                                                                   \\ \midrule
			ByteVector                          & 81                              & 64.29\%                               & 20.24\%                                 & 15.48\%                                    \\
			HSLColor                            & 96                              & 54.44\%                               & 42.22\%                                 & 3.33\%                                     \\ \bottomrule
		\end{tabular}
	}
\end{table}

Given the classification of equivalent and non-equivalent mutants
allows us to look at player behavior, i.e., how many equivalent
mutants are usually created in a game of \toolname.
\Cref{tab:equimutants} shows the average ratio of equivalent mutants
among all submitted mutants, suggesting a range of 2\% to
approximately 13\% per game with a total of 5.75\%.

Notably, the \texttt{Options} CUT exhibits the lowest share of equivalent mutants, aligning with its highest percentage of mutants detected as not equivalent and the lowest percentage of mutants automatically identified as equivalent. Conversely, the \texttt{ByteVector} and \texttt{HSLColor} CUTs have the highest ratio of equivalent mutants. Interestingly, games for these two CUTs utilized the \textit{Intention Collection} feature, where players indicated whether they intended to create an equivalent mutant or not. With this feature enabled, players might intentionally create more equivalent mutants because they are aware of the possibility. Conversely, in games featuring other CUTs, people may create most of their equivalent mutants unintentionally.

While manually inspecting the sample of remaining unknown mutants, we observed that some mutants were trivially equivalent. These mutants often employed patterns such as adding equivalent arithmetic (e.g., adding \texttt{+<value>-<value>} with the value typically being \texttt{+0} and \texttt{*1} like in \cref{lst:triv1}), introducing unnecessary calls to methods or field declarations as depicted in \cref{lst:trv2}, or expanding comparisons which would result in the same return value (see \cref{lst:trv3}). We conjecture that these are intentionally created equivalent mutants: If playing in \emph{hard} mode the actual syntactical change is not shown to defenders, and in that case, it does not matter what an equivalent mutant looks like.

There are also non-trivial equivalent mutants, such as shown in \cref{lst:nontriv}: The mutant wraps a method call, computing the absolute value around another method that retrieves the index of a key, which is always a positive number; this mutant would also be the result of a traditional ``absolute value insertion'' mutation, but it can be detected neither by TCE nor by TCE+.

\begin{lstlisting}[float, language=diff, float, caption=Non-trivial equivalent mutant in \texttt{Options}, captionpos=b, label=lst:nontriv, linewidth=\columnwidth]
   if (requiredOpts.contains(key)) {
-    requiredOpts.remove(
-      requiredOpts.indexOf(key));
+    requiredOpts.remove(
+      Math.abs(requiredOpts.indexOf(key)));
   }
\end{lstlisting}

\begin{lstlisting}[float, language=diff, float, caption=Trivial equivalent mutant in \texttt{Complex} not found bei TCE(+), captionpos=b, label=lst:triv1, linewidth=\columnwidth]
   public Complex pow(double power) {
-    double r = abs();
+    double r = abs()+0.0;
     double theta = angle();
\end{lstlisting}
	
\begin{lstlisting}[float, language=diff, float, caption=Trivial equivalent mutant in \texttt{Document} not found bei TCE(+), captionpos=b, label=lst:trv2, linewidth=\columnwidth]
   @Override
   public Iterator<IndexableField> iterator() {
+    fields.toString();
     return fields.iterator();
   }
\end{lstlisting}
	
\begin{lstlisting}[float, language=diff, float, caption=Trivial equivalent mutant in \texttt{HSLColor} not found bei TCE(+), captionpos=b, label=lst:trv3, linewidth=\columnwidth]
   private int iMax(int a, int b) {
-    if (a > b) return a; else return b;
+    if (a >= b) return a; else return b;
   }
\end{lstlisting}

\begin{lstlisting}[float, language=diff, float, caption=Mutant difficult to detect in \texttt{Options}, captionpos=b, label=lst:diff1, linewidth=\columnwidth]
   if (longOpts.keySet().contains(opt)) {
-    return Collections.singletonList(opt);
+    List<String> list = new ArrayList<>();
+    list.add(opt);
+    return list;
   }
\end{lstlisting}

\begin{lstlisting}[float, language=diff, float, caption=Mutant difficult to detect in \texttt{Options}, captionpos=b, label=lst:diff2, linewidth=\columnwidth]
-  if (str.startsWith("--")) {
+  if (str.contains("--")) {
     return str.substring(2);
   }
	\end{lstlisting}

Some mutants in our dataset also proved challenging to detect rather than equivalent. For example, \cref{lst:diff1} shows a mutant that replaces the return type from a \texttt{SingletonList} to an \texttt{ArrayList}, a change that can only be identified by inspecting the type using the \texttt{instanceOf} operator. This mutant is hard to detect, but  may not be desirable for mutation testing since bugs in return types would be caught by the compiler. On the other hand, \cref{lst:diff2} shows a mutant that can only be detected if an incorrect \texttt{String} input is used, containing a double hyphen somewhere within. This appears to be a strong and useful mutant since it can reveal potentially serious bugs in \texttt{String} manipulation.

\summary{RQ 2}{In total 5.75\% of all player-submitted mutants are equivalent, which is 6.94\% per CUT on average.}

\subsection{RQ 3: How well do players perform at detecting (non-) equivalent mutants?}

\begin{table}[t]
	\caption{Ratio of equivalent mutants (EM) in duels in \%}
	\label{tab:equmutantduels}
	\centering
	%	\resizebox{\linewidth}{!}{%
	\begin{tabular}{lrrr}
		\toprule
		\textbf{CUT Alias} & \textbf{Overall} & \textbf{Manual duels} & \textbf{Automatic duels} \\ \midrule
		ByteVector         & 42.42                      & 37.74                             & 61.54                                \\
		Complex\_V1        & 36.11                      & 38.46                             & 40.00                                \\
		Complex\_V2        & 27.12                      & 27.12                             & --.--                                \\
		Document           & 20.92                      & 19.84                             & 28.57                                \\
		HSLColor           & 30.00                      & 28.24                             & 29.52                                \\
		IntHashMap         & 7.31                       & 6.27                              & 8.72                                 \\
		Lift               & 33.80                      & 35.84                             & 37.50                                \\
		Options            & 5.61                       & 6.29                              & 8.11                                 \\
		Rational           & 50.00                      & 50.00                             & --.--                                \\
		SparseIntArray     & 20.92                      & 24.28                             & 12.08                                \\ \midrule
		Average			   & 27.42						& 27.41								& 28.26								   \\ \bottomrule
	\end{tabular}
	%	}
\end{table}

Of those mutants with involved in equivalent duels, the proportion of
equivalent mutants was always 50\% or lower
(\cref{tab:equmutantduels}), which suggests that defenders frequently
gave up, or overestimated the quality of their tests.

On two CUTs (\texttt{Complex\_V2} and \texttt{Rational}) the option to trigger equivalence duels automatically was disabled and therefore no automatic duels were triggered. For all other CUTs, the ratio of equivalent mutants is almost always higher for automatically triggered duels than for manual ones (\cref{tab:equmutantduels}). Defenders, restricted to seeing only the line where the mutant is located rather than the mutated code itself, may use many attempts to reveal a mutant, which may be somewhat unfair if that mutant is equivalent and the defenders are persistent. The higher ratio of equivalent mutants suggests that automatically triggering an equivalence duel after several attempts (10 by default) achieves its purpose of reducing such futile attempts, thus ensuring continuing gameplay.

%This approach could inadvertently trigger an automated duel, as the default threshold is set to ten tests covering the mutant.
%Conversely, lacking the knowledge or skill to conduct more targeted testing on specific lines, they may deem the mutant equivalent simply to eliminate it.

When submitting a test in a duel, there are two potential outcomes:
either the test successfully eliminates the mutant or the mutant
survives the tests. About 45\% of duels where attackers submitted a
test resulted in the successful elimination of the mutant, as shown in
\cref{tab:duels}. The remaining 55\% of mutants survived, either being
equivalent or not. Among these surviving mutants, 57\% were equivalent
(37.35\% of all duels resolved with a test), but the attackers failed
to recognize that and tried to write a killing test
anyway. Conversely, 43\% were not equivalent (23.65\% of all duels
resolved with a test), yet the players still failed to write a killing
test.

When attackers assume a mutant is equivalent, then during a duel they
can indicate this by selecting ``accept mutant as equivalent''.
\Cref{tab:duels} illustrates the number of correctly and incorrectly
accepted equivalent mutants, revealing that only around 35\% were
accurately identified as equivalent, while approximately 65\% were
mutants that were not equivalent. This again indicates that the
attackers were not proficient at identifying equivalent mutants, or
perhaps they were simply eager to conclude the duel swiftly to resume
mutant creation.

\Cref{tab:mutantintention} displays the number of mutants and their proportions categorized based on whether their intention information was correct, incorrect, or not provided by the player. Notably, 88\% of mutants were correctly classified for both classes, but \texttt{HSLColor} had a slightly higher proportion of incorrectly classified mutants compared to \texttt{ByteVector}, which had more unclassified mutants. Focusing on non-equivalent mutants within the dataset of mutants with intentions (\cref{tab:mutantintention}), players demonstrated proficiency with over 92\% being correctly classified as not equivalent, with an error rate below 2\% for both CUTs. However, for equivalent mutants within the dataset of mutants with intentions, the ratio of correct classified intentions by the attackers is 64\% and 54\% for \texttt{ByteVector} and \texttt{HSLColor}, respectively, indicating a lower accuracy than the overall correctness ratio suggests, and generally more uncertainty about equivalent mutants.

\summary{RQ3}{While the majority of players accurately indicated their intention to create an equivalent mutant or not, nearly two-thirds of them were unable to correctly identify equivalent mutants created by other players or themselves.}
%%% Local Variables:
%%% mode: latex
%%% TeX-master: "../main"
%%% End:

\section{Related Work}

Several papers have explored Code Defenders, focusing on mutants and tests. An analysis of 20 games, each featuring a different CUT, reports an average mutation score of 69.48\% but does not delve into equivalent mutants~\cite{DBLP:conf/icse/RojasWCF17}. A further study of 12 classes with multiple games per class, showed that 85\% of valid mutants could be detected as killable~\cite{DBLP:conf/sigcse/FraserGKR19}, but provided only a surface-level overview of player interactions with equivalence duels. Our study builds on these findings by leveraging tests from multiple years to identify killable mutants, revealing that over 91\% of valid mutants, on average, can be detected by these tests. Additionally, a thorough examination of equivalent mutants is conducted, estimating the ratio of equivalent mutants per CUT through automatic equivalence detection and manual investigation of a randomly sampled subset of remaining unknown mutants.

For automatic equivalent mutant detection, we used TCE~\cite{DBLP:conf/icse/PapadakisJHT15} and TCE+~\cite{DBLP:conf/fsen/HoushmandP17}. TCE, originally designed for C programs, underwent evaluation on a tagged mutant dataset, while TCE+ was evaluated on automatically generated and expert-tagged mutants. In contrast, this paper employs manually created human mutants from Code Defenders, not tagging all mutants but using existing tests to exclude known killable mutants. TCE was found incapable of detecting Java mutants and reported TCE+ detecting 18\% to 100\% of equivalent mutants~\cite{DBLP:conf/fsen/HoushmandP17}. In this study, TCE+ is more effective, but TCE is also capable of detecting equivalent mutants. However, TCE+ does not detect over 70\% of equivalent mutants for any CUT in this dataset.

People's proficiency in classifying mutants as equivalent or not was
initially examined using four Cobol programs with roughly 40\%
equivalent mutants~\cite{acree1980mutation}. Competent programmers
correctly classified 80\% of mutants, misclassifying 12\% killable
mutants and 33\% equivalent mutants. In contrast, our intention data
involves two programs with a larger number of people who only
classified their own mutants.
%The ratio of killable to
%equivalent mutants is more substantial, and players had the option to
%express uncertainty.
Results from this experiment show differences but align with a similar
trend: players classified 88\% of mutants correctly while
misclassifying 8\% of killable and 40\% of equivalent mutants.

We studied mutants created manually in \toolname.  Deep learning
models have been recently suggested to create mutants that resemble
real
faults~\cite{DBLP:conf/icsm/TufanoWBPWP19,DBLP:conf/icse/TufanoKWWBPP20,degiovanni2022mubert}. It
is conceivable that mutants created by deep learning models resemble
human-written equivalent mutants more than those created by
traditional operators, but further research is required.

%In a different approach, a model was trained to predict whether a mutant, when subjected to a given test suite, would be killed or survive without actual execution~\cite{DBLP:journals/access/NaeemLNUS19}. While this method proved effective, it does not distinguish between equivalent and non-equivalent surviving mutants. Furthermore, there is a proposal for a neural network designed to differentiate between equivalent and non-equivalent surviving mutants~\cite{jammalamadaka2022equivalent}. However, it is important to note that this approach remains theoretical, with only simulation results indicating its functionality at this stage.
%%% Local Variables:
%%% mode: latex
%%% TeX-master: "../main"
%%% End:

\section{Conclusions}

Up to 13\% of mutants created by humans are equivalent, some
intentionally crafted and others created by accident. A substantial
share of these equivalent mutants can be found using TCE+, which is
important since human classification of equivalent mutants is prone to
errors, with incorrect identification occurring in almost two-thirds
of cases.
%,
%highlighting the potential for manual classification to introduce
%significant errors.

Expanding the capabilities of \toolname presents opportunities to
enhance players' understanding of equivalent mutants. One approach
might be to incorporate puzzles specifically designed for resolving
equivalence duels, educating players on how to identify equivalent
mutants. Additionally, an option could be introduced in equivalence
duels where players can indicate that a mutant is not equivalent, but
they lack the knowledge to write a test to prove it. A more
comprehensive study involving larger projects could yield deeper
insights into human proficiency in detecting equivalent mutants.

In order to support experiment replications and further research
on mutation testing, our dataset is available at:

\begin{center}
	\begin{center}
		\url{https://doi.org/10.6084/m9.figshare.25144313}
	\end{center}
\end{center}

%%% Local Variables:
%%% mode: latex
%%% TeX-master: "../main"
%%% End:

\balance
\bibliographystyle{ieeetr}
\bibliography{bib}

\begin{thebibliography}{10}

\bibitem{DBLP:journals/tse/PetrovicIFJ22}
G.~Petrovic, M.~Ivankovic, G.~Fraser, and R.~Just, ``Practical mutation testing
  at scale: {A} view from google,'' {\em {IEEE} Trans. Software Eng.}, vol.~48,
  no.~10, pp.~3900--3912, 2022.

\bibitem{DBLP:journals/computer/DeMilloLS78}
R.~A. DeMillo, R.~J. Lipton, and F.~G. Sayward, ``Hints on test data selection:
  Help for the practicing programmer,'' {\em Computer}, vol.~11, no.~4,
  pp.~34--41, 1978.

\bibitem{DBLP:journals/stvr/OffuttC94}
A.~J. Offutt and W.~M. Craft, ``Using compiler optimization techniques to
  detect equivalent mutants,'' {\em Softw. Test. Verification Reliab.}, vol.~4,
  no.~3, pp.~131--154, 1994.

\bibitem{DBLP:conf/icse/PapadakisJHT15}
M.~Papadakis, Y.~Jia, M.~Harman, and Y.~L. Traon, ``Trivial compiler
  equivalence: {A} large scale empirical study of a simple, fast and effective
  equivalent mutant detection technique,'' in {\em 37th {IEEE/ACM}
  International Conference on Software Engineering, {ICSE} 2015}, pp.~936--946,
  {IEEE} Computer Society, 2015.

\bibitem{DBLP:conf/fsen/HoushmandP17}
M.~Houshmand and S.~Paydar, ``{TCE+:} an extension of the {TCE} method for
  detecting equivalent mutants in java programs,'' in {\em Fundamentals of
  Software Engineering - 7th International Conference, {FSEN} 2017}, vol.~10522
  of {\em LNCS}, pp.~164--179, Springer, 2017.

\bibitem{naeem2020machine}
M.~R. Naeem, T.~Lin, H.~Naeem, and H.~Liu, ``A machine learning approach for
  classification of equivalent mutants,'' {\em Journal of Software: Evolution
  and Process}, vol.~32, no.~5, p.~e2238, 2020.

\bibitem{DBLP:journals/stvr/SchulerZ13}
D.~Schuler and A.~Zeller, ``Covering and uncovering equivalent mutants,'' {\em
  Softw. Test. Verification Reliab.}, vol.~23, no.~5, pp.~353--374, 2013.

\bibitem{DBLP:journals/jss/MarsitAKLLOM21}
I.~Marsit, A.~Ayad, D.~Kim, M.~Latif, J.~M. Loh, M.~N. Omri, and A.~Mili, ``The
  ratio of equivalent mutants: {A} key to analyzing mutation equivalence,''
  {\em J. Syst. Softw.}, vol.~181, p.~111039, 2021.

\bibitem{DBLP:conf/icst/Pitts22}
R.~Pitts, ``Random mutant selection and equivalent mutants revisited,'' in {\em
  {IEEE} International Conference on Software Testing, Verification and
  Validation Workshops {ICST} Workshops 2022}, pp.~170--178, {IEEE}, 2022.

\bibitem{DBLP:conf/icst/RojasF16}
J.~M. Rojas and G.~Fraser, ``Code defenders: {A} mutation testing game,'' in
  {\em Int. Conference on Software Testing, Verification and Validation
  Workshops}, pp.~162--167, {IEEE} Computer Society, 2016.

\bibitem{DBLP:conf/sigcse/FraserGKR19}
G.~Fraser, A.~Gambi, M.~Kreis, and J.~M. Rojas, ``Gamifying a software testing
  course with code defenders,'' in {\em {ACM} Technical Symposium on Computer
  Science Education, {SIGCSE} 2019}, pp.~571--577, {ACM}, 2019.

\bibitem{DBLP:conf/icse/RojasWCF17}
J.~M. Rojas, T.~D. White, B.~S. Clegg, and G.~Fraser, ``Code defenders:
  crowdsourcing effective tests and subtle mutants with a mutation testing
  game,'' in {\em Proceedings of the 39th International Conference on Software
  Engineering, {ICSE} 2017}, pp.~677--688, {IEEE} / {ACM}, 2017.

\bibitem{DBLP:conf/icsm/TufanoWBPWP19}
M.~Tufano, C.~Watson, G.~Bavota, M.~D. Penta, M.~White, and D.~Poshyvanyk,
  ``Learning how to mutate source code from bug-fixes,'' in {\em 2019 {IEEE}
  International Conference on Software Maintenance and Evolution, {ICSME}
  2019}, pp.~301--312, {IEEE}, 2019.

\bibitem{DBLP:conf/icse/TufanoKWWBPP20}
M.~Tufano, J.~Kimko, S.~Wang, C.~Watson, G.~Bavota, M.~D. Penta, and
  D.~Poshyvanyk, ``Deepmutation: a neural mutation tool,'' in {\em
  International Conference on Software Engineering}, pp.~29--32, {ACM}, 2020.

\bibitem{degiovanni2022mubert}
R.~Degiovanni and M.~Papadakis, ``$\mu$bert: Mutation testing using pre-trained
  language models,'' in {\em Int. Conf. on Software Testing, Verification and
  Validation Workshops (ICSTW)}, pp.~160--169, IEEE, 2022.

\bibitem{DBLP:conf/afips/BuddLDS78}
T.~A. Budd, R.~J. Lipton, R.~A. DeMillo, and F.~G. Sayward, ``The design of a
  prototype mutation system for program testing,'' in {\em American Federation
  of Information Processing Societies: 1978 National Computer Conference},
  vol.~47, pp.~623--629, {AFIPS} Press, 1978.

\bibitem{walsh1985measure}
P.~J. Walsh, {\em A measure of test case completeness (software, engineering)}.
\newblock State University of New York at Binghamton, 1985.

\bibitem{DBLP:journals/acta/BuddA82}
T.~A. Budd and D.~Angluin, ``Two notions of correctness and their relation to
  testing,'' {\em Acta Informatica}, vol.~18, pp.~31--45, 1982.

\bibitem{yao2014study}
X.~Yao, M.~Harman, and Y.~Jia, ``A study of equivalent and stubborn mutation
  operators using human analysis of equivalence,'' in {\em Int. Conference on
  Software Engineering (ICSE)}, pp.~919--930, 2014.

\bibitem{zhang2016predictive}
J.~Zhang, Z.~Wang, L.~Zhang, D.~Hao, L.~Zang, S.~Cheng, and L.~Zhang,
  ``Predictive mutation testing,'' in {\em Proceedings of the 25th
  International Symposium on Software Testing and Analysis}, pp.~342--353,
  2016.

\bibitem{peacock2021automatic}
S.~Peacock, L.~Deng, J.~Dehlinger, and S.~Chakraborty, ``Automatic equivalent
  mutants classification using abstract syntax tree neural networks,'' in {\em
  2021 IEEE International Conference on Software Testing, Verification and
  Validation Workshops (ICSTW)}, pp.~13--18, IEEE, 2021.

\bibitem{brito2020preliminary}
C.~Brito, V.~H. Durelli, R.~S. Durelli, S.~R. de~Souza, A.~M. Vincenzi, and
  M.~E. Delamaro, ``A preliminary investigation into using machine learning
  algorithms to identify minimal and equivalent mutants,'' in {\em 2020 IEEE
  International Conference on Software Testing, Verification and Validation
  Workshops (ICSTW)}, pp.~304--313, IEEE, 2020.

\bibitem{jammalamadaka2022equivalent}
K.~Jammalamadaka and N.~Parveen, ``Equivalent mutant identification using
  hybrid wavelet convolutional rain optimization,'' {\em Software: Practice and
  Experience}, vol.~52, no.~2, pp.~576--593, 2022.

\bibitem{van2021mutantbench}
L.~van Hijfte and A.~Oprescu, ``Mutantbench: an equivalent mutant problem
  comparison framework,'' in {\em Int. Conference on Software esting,
  Verification and Validation Workshops (ICSTW)}, pp.~7--12, IEEE, 2021.

\bibitem{chung2022augmenting}
S.~Chung and S.~Yoo, ``Augmenting equivalent mutant dataset using symbolic
  execution,'' in {\em Int. Conference on Software Testing, Verification and
  Validation Workshops (ICSTW)}, pp.~150--159, IEEE, 2022.

\bibitem{DBLP:conf/icse/ElbaumPDJ07}
S.~G. Elbaum, S.~Person, J.~Dokulil, and M.~Jorde, ``Bug hunt: Making early
  software testing lessons engaging and affordable,'' in {\em 29th
  International Conference on Software Engineering ({ICSE} 2007)},
  pp.~688--697, {IEEE} Computer Society, 2007.

\bibitem{DBLP:conf/fie/OliveiraOCD15}
R.~A.~P. Oliveira, L.~B.~R. Oliveira, B.~B.~P. Cafeo, and V.~H.~S. Durelli,
  ``Evaluation and assessment of effects on exploring mutation testing in
  programming courses,'' in {\em 2015 {IEEE} Frontiers in Education Conference,
  {FIE} 2015}, pp.~1--9, {IEEE} Computer Society, 2015.

\bibitem{acree1980mutation}
A.~T. Acree~Jr, {\em On mutation}.
\newblock Georgia Institute of Technology, 1980.

\end{thebibliography}

\end{document}